\documentstyle[mprocl,psfig]{article}

\def\NPB{{\em Nucl. Phys.} B}
\def\PLB{{\em Phys. Lett.}  B}
\def\PRL{\em Phys. Rev. Lett.}
\def\PRD{{\em Phys. Rev.} D}
\def\ZPC{{\em Z. Phys.} C}

\def\be{\begin{equation}}
\def\ee{\end{equation}}
\def\bea{\begin{eqnarray}}
\def\eea{\end{eqnarray}}

\def\step{\\[-.25cm]}

\def\ov{\over}

\date{\today}
\let\Ga=\Gamma

\def\ds{\displaystyle}

\def\dcal{{\cal D}}

\def\tr{{\rm tr}}
\def\Tr{{\rm Tr}}
\def\Der{{\bf {\cal{G}}}}
\def\Loop{{L}}
\def\eq#1{(\ref{#1})}

\def\di{\displaystyle}
\def\longbar#1{#1\kern-0.7em\raise1.3ex\hbox{{$-$}}}
\def\tab{&\di}
\let\de=\delta

\begin{document}
 \begin{flushright}
 {\tt DIAS-STP-01-01\\
 CERN-TH-2001-010\\
 FAU-TP3-01-01}
 \end{flushright}
\vspace{0cm}
\title{GAUGE INVARIANCE, BACKGROUND FIELDS\\AND MODIFIED WARD
  IDENTITIES}

\author{F. FREIRE}

\address{School of Theoretical Physics,
       Dublin Institute for Advanced Studies,\\
       10 Burlington Road, Dublin 4, Ireland}

\author{D. F. LITIM}

\address{Theory Division, CERN,\\
CH-1211 Geneva 23, Switzerland}

\author{J. M. PAWLOWSKI}

\address{Institut f\"ur Theoretische Physik III,
Universit\"at Erlangen-N\"urnberg,\\Staudtstra\ss e 7,
D-91058 Erlangen, Germany}


\maketitle\abstracts{In this talk the gauge symmetry for Wilsonian
  flows in pure Yang-Mills theories is discussed. The background field
  formalism is used for the construction of a gauge invariant
  effective action. The symmetries of the effective action
  under gauge transformations for both the gauge field and the
  auxiliary background field are separately evaluated. Modified
  Ward\,-Takahashi and background field identities are used in my study.
  Finally it is shown how the symmetry properties of the full theory
  are restored in the limit where the cut-off is removed.}
\vspace{-.05cm}
\noindent
{\sf Introduction and aim\,:}
Wilsonian or Exact renormalisation group (ERG) equations have
successfully been applied to non-perturbative phenomena in quantum
field theories. Hence an ERG formulation of gauge theories is a
promising tool for resolving open questions concerning the 
non-perturbative regime of these theories, e.g.\ confinement, 
chiral symmetry breaking.  A key hurdle in such a task concerns 
the consistent and practicable introduction of an infra-red cut-off in
a theory with a non-linear local symmetry. 

In this talk\footnote{talk presented by FF at the Second Conference on
  the\\Exact Renormalization Group, Roma, Sep.\,\,18-22, 2000.} I will discuss
this difficult matter in some detail.  The quest can be presented as follow\,:
How can it be ensured that a Wilsonian effective action
shows the gauge symmetry of the underlying full theory\,? Let me first 
explain in more detail why this question requires a surgical look into.    
Implementations of the ERG\,\cite{Polchinski:1984gv} are most intricate
when the symmetries of the theory are deformed by the intrinsic
infra-red cut-off to this approach. As the integration of 
the flow equation is carried out it is necessary to guarantee that
the information about the inherent symmetries of the theory
is not washed out. In general, much work has
been devoted to overcoming the involved problems mainly within 
different approaches to non-Abelian theories.\,\cite{Reuter:1994kw,Ellwanger:1994iz,Morris:2000px,Litim:1998nf} 
Here, for the sake of clarity, I will discuss 
this problem for pure Yang-Mills theory. Moreover I use an approach 
to these theories within the background field formalism which permits the 
definition of a gauge invariant effective action. 

The aim is to clarify how it can be ensured that after the
complete integration of the ERG equations, {\it i.e.\/} flow the cut-off scale
$k$ to zero, a gauge invariant solution is obtained without the
need of an extra fine-tuning.  Furthermore, I will
show that the resurgent symmetry does indeed correspond to the inherent gauge
one. Most technical details will be bypassed in this presentation
and may be found in a recent letter.\,\cite{Freire:2000bq}

At the centre of my presentation is the quest of understanding how
physical information is encrypted along the flow through an
interplay between gauge invariance, Ward\,-Takahashi identities and
background field identities. A similar programme has been pursued a 
few years ago in the context of Abelian 
theories.\,\cite{Freire:1996db}
Background field identities were shown to contain the Ward identities
under a requirement of gauge invariance.

The key ingredient of the present approach is the ERG equation for the
effective action $\Gamma_k$. It describes 
the logarithmic rate of the change of $\Gamma_k$
with respect to the scale $k$.  Following the standard implementation
of the background field formalism I introduce a non-dynamical auxiliary field
$\bar A$, the so-called background gauge field. Then
formally the flow equation for pure Yang-Mills may be written as\,:
\begin{eqnarray}
\partial_t\Gamma_k[A,c,c^*;\bar A] =
\frac{1}{2}\Tr\left(\frac{\de^2\Gamma_k}{\delta A\delta A}
              +R_{\!A}\right)^{\!\!\!-1}\!\!\!\!\partial_t  R_{\!A}
\,- \,
\Tr \left( \frac{\de^2\Gamma_k}{\delta c \delta c^*} 
    +R_{C}\right)^{\!\!\!-1}\!\!\!\! \partial_t
              R_{C}\,.
\label{flow}
\end{eqnarray}
I use the common notation where $t=\ln k$ and the trace $\Tr$ 
denotes a sum over momenta, Lorentz and gauge group indices. 
The functions $R_{\!A}$ and $R_{C}$ implement the
infra-red cut-off for the gauge field $A$ and
ghost fields $c$ and $c^*$ respectively. They may also depend on the
background field, to which I now turn your attention.\step

\noindent
{\sf Background field formalism\,:}
I briefly summarise some important points about the background
field formalism, in particular the r\^ole of different
gauge transformations. The formalism is settled
on the use of a background field dependent gauge-fixing condition that is 
invariant under a simultaneous gauge transformation of $\bar A$ 
and of the fields $A, c$ and $c^*$. This can be used for 
a definition of an effective action which is invariant under 
this combined gauge transformation. 
As $\bar A$ is involved in this transformation, the invariance of the
effective action is, {\it a priori}, only an auxiliary symmetry.
However, for the choice $\bar A = A$ it becomes the inherent gauge symmetry 
of the theory.

For a pure Yang-Mills theory including the ghost term,
\begin{eqnarray}\label{S} 
S=S_{\rm cl} + S_{\rm gf} + S_{\rm gh}\,.
\end{eqnarray} 
The classical action $S_{\rm cl}=\frac{1}{4}\int_x
F^a_{\mu\nu}F^a_{\mu\nu}$ contains the field strength tensor
$F_{\mu\nu}(A) = \partial_{\mu} A_\mu -\partial_\nu A_{\mu} +
g\,[A_{\mu},\,A_\nu]$, where $F_{\mu\nu} \equiv F^a_{\mu\nu}t^a$ and
$A_{\mu} = A^a_{\mu} t^a$ with the generators $t^a$
satisfying $[t^a,\,t^b]=f^{abc}t^c$ and $\tr\, t^a t^b 
=-\frac{1}{2}\,\de^{ab}$. I also use the
shorthand notation $\int_x \equiv \int d^dx$. In the adjoint
representation, the covariant derivative is  
\begin{eqnarray}\label{covD}
D_\mu^{ab}(A) = \delta^{ab}\partial_\mu + g f^{acb}A^c_\mu\ .  
\end{eqnarray} 
The natural choice for the gauge fixing is the so-called background
field gauge. The corresponding gauge-fixing and ghost actions are
respectively,
\begin{equation}\label{gauge} 
S_{\rm gf} =-\frac{1}{2\xi}\int_x\ 
(A-\bar A)^a_{\mu}\,\longbar D{}_\mu^{ab} \longbar D{}_\nu^{bc}\,
(A-\bar A)^c_{\nu}\ , \qquad 
S_{\rm gh} =-\int_x c^*_a \,\longbar D{}_\mu^{ac}D_\mu^{cd}\,c_d\ ,
\end{equation}
which involves the covariant derivative $\longbar D{}\equiv D(\bar A)$. 
The symmetries of the action in \eq{S}
can be inspected by two different gauge
transformations. The first one gauge transforms the dynamic fields
$A,\,c,\,\textrm{and}\,c^*$\,-- it represents the underlying symmetry 
of the theory. 
Its infinitesimal generator $\Der^a$ in a natural representation is
defined as
\begin{eqnarray}\label{delkernel} 
\Der^a\tab = \tab D_\mu^{ab}\frac{\delta}{\delta A_\mu^b} -
g\,f^{abc}\left( c_c\, \frac{\delta }{\delta c_b}+
  c^*_c\,\frac{\delta }{\delta c^*_b}\right)\ .
\end{eqnarray}
From the action of $\Der^a$
on the fields it can be shown that $A$ transforms inhomogeneously, the
ghosts transform as tensors and $\bar A$ is invariant. It follows that
the covariant derivative transforms as a tensor.

The second transformation, denoted by the generated
$\bar\Der{}^a$, acts only on the background field,
\begin{eqnarray}\label{delbarkernel} 
\bar\Der^a\tab =\tab \longbar D{}_\mu^{ab}\frac{\delta}{\delta \bar
  A_\mu^b}\ ,
\end{eqnarray}
and under its action $\bar A$ transforms inhomogeneously like $A$ under
$\Der^a$, and therefore the covariant derivative $\longbar D$ also
transforms as a tensor. Note that the auxiliary transformation
$\bar\Der^a$ as it stands, does not carry any physical information.

I now turn your attention to the manner in which $\Der^a$ and
$\bar\Der^a$ operate on the action $S$.
The classical action is invariant under both transformations
since it does not depend on the background field,
while neither $S_{{\rm gf}}$ nor
$S_{{\rm gh}}$ are invariant under \eq{delkernel} or \eq{delbarkernel}. 
For \eq{delkernel} it follows,
\begin{eqnarray}\label{vary1} 
\Der^a(x) S_{\rm gf} =  
{1\ov \xi} D^{ab}_\mu\longbar D{}^{bc}_\mu\longbar D{}^{cd}_\nu 
(A-\bar A)^d_\nu(x)\ ,  \quad \tab  
\Der^a(x)  S_{\rm gh} =  f^{bdc}\longbar D{}_{\mu}^{ad}\left( c^*_b 
D^{ce}_\mu c_e\right)\ .   
\end{eqnarray} 
Now from the explicit expression for $S_{\rm gf}$ and
$S_{\rm gh}$, as given in \eq{gauge}, 
it follows that \eq{vary1} just displays 
$-\bar\Der^a S_{\rm gf}$ and $-\bar\Der^a  S_{\rm gh}$ respectively. 
Thus, in the background
field gauge, each term in the Yang-Mills action~(\ref{S}) 
is {\it separately} invariant under the combined transformation
$\Der+\bar\Der$. A key point of the
background field formalism has been reached\,:
the action resulting from setting the background field
equal to the gauge field, \textit{i.e.}\
$\hat S[A,c,c^*]\!\equiv S[A,c,c^*;\bar A\!=\!A]$
is invariant under the {\it physical} symmetry generated by \eq{delkernel},
$\Der^a \hat S[A,c,c^*]=0$, with $S[A,c,c^*;\bar A]$ 
satisfying the classical `Ward\,-Takahashi identity', $\Der^a S=\Der^a 
(S_{\rm gf}+S_{\rm gh})$. 

At quantum level this symmetry turns into the
gauge invariance of the effective action $\Gamma[A,c,c^*;\bar
A\!=\!A]$, which in turn satisfies the Ward\,-Takahashi identity for a 
non-Abelian gauge theory.  However I remind you that it is only the
combination of both statements that gives a physical meaning to this
gauge invariance. Note that heuristically this result stems from the
observation that in the quantised theory the source only couple to the
fluctuation field
\begin{eqnarray}
a^a_\mu = A^a_\mu -\bar A^a_\mu\ ,  
\end{eqnarray} 
and the gauge fixing condition~\eq{gauge} only constrains
$a^a_\mu$.\step

\noindent
{\sf Background field dependent Wilsonian flows\,:}
In order to effectively implement the background field formalism for
the coarse-grained effective Yang-Mills action I choose the regulator terms
\begin{eqnarray}\label{Sk} 
\Delta S_{k}=\Delta S_{k,A}+ \Delta
S_{k,C}\ ,
\end{eqnarray}
for the gauge and the ghost fields, respectively, to be\,\cite{Reuter:1994kw}
\begin{eqnarray}        \label{DeltaSA} 
\Delta S_{k,A} &=& \frac{1}{2}
\int_x (A-\bar A)^a_{\mu}\,R_{\!A}{}^{ab}_{\mu\nu}(P_{\!\!\!A}^2)\,
(A-\bar A)^b_{\nu}  
\\                      \label{DeltaSC} 
\Delta S_{k,C} &=&  \int_x c^*_a \,R_{C}^{ab}(P_{\!\!C}^2)\,c_b\ .
\end{eqnarray}
The arguments $P_{\!\!\!A}^2$ and $P_{\!\!C}^2$ of the
regulator functions are appropriately defined background field
dependent Laplaceans, and their choice might determine the symmetries
of the resulting theory. For instance, in order to have an 
action $S_{k}$ which is gauge invariant under the combined
transformation it is only necessary 
to required that both $P_{\!\!\!A}^2$ and $P_{\!\!C}^2$ transform as
tensors under $\Der+\bar\Der$. Thus, in the following,
I shall assume that such a choice has been made. 

Up to this point, I have restricted the presentation  to the classical action 
with the regulator terms \eq{Sk} added. The computation of the
coarse-grained effective action $\Gamma_k$ follows the usual
procedure. Consider the Schwinger functional
$W_k \equiv W_k[J_{\mu},\,\eta,\,\eta^*;\, \bar A_\mu]$,
\begin{eqnarray}\label{Schwinger}
  \!\!\exp W_k =\!\ds\int\prod_a\left\{\dcal A^a_{\mu}\,\dcal c_a\,\dcal
  c^*_a\right\}\,\exp\left[-S_k
  +\!\int(J^a_{\mu}(A\!-\!\bar A)^a_{\mu}+\eta^*_a c_a\!-\!c^*_a\eta_a)\right]\!.
\end{eqnarray}\step
where $(J,\eta,\eta^*)$ are the respective sources.
Then the effective action $\Gamma_{k}$ is given by
\begin{eqnarray}\nonumber
\Gamma_k[A,c,c^*; \bar A] =
- W_k[J,\eta,\eta^*;\bar A] &-& \Delta S_k[A,c,c^*;\bar A]\\
&+& \int_{x}\!\!\left(J^a_{\mu}(A -\!\bar A)^a_{\mu} + \eta^*_a
\bar c_a - c^*_a\eta_a\right). \label{defofG}
\end{eqnarray}\step
The flow equation for $\Gamma_k$ has already been given at the
beginning, \eq{flow}. Now I introduce the effective action $\hat
\Gamma_k$,
\begin{equation}
\hat\Gamma_k[A,c,c^*] \equiv\Gamma_k[A,c,c^*;\, \bar A=A]. 
\end{equation} 
As I shall argue later, this new action is gauge-invariant.
Its flow equation, of course, is given by the
flow of $\Gamma_k$ in \eq{flow}, but evaluated at $\bar A=A$. 
It is important to stress that $\partial_t \hat \Gamma_k$, since it
depends on the second functional derivatives of $\Gamma_k$ with
respect to the dynamical fields (at $\bar A=A$), is
a functional of $\Gamma_k$ and {\it not} a functional of $\hat
\Gamma_k$. This means that is not sufficient to study the
symmetries of $\hat\Gamma_k$ but also necessary to study those of
$\Gamma_k$.\step

\noindent
{\sf Modified and background field Ward\,-Takahashi identities\,:}
I will now discuss the Ward\,-Takahashi identities
that are related to the gauge transformations \eq{delkernel}
and \eq{delbarkernel}.
In the Wilsonian formalism, due to the presence of a coarse-graining,
these identities receive a contribution from the regulator term.  
The identity that follows from considering
$\Der^a\Gamma_k$ is called the {\it modified} Ward\,-Takahashi
identities (mWI).  A second identity follows from $\bar\Der^a\Gamma_k$
and I shall denote it as 
the {\it background field} Ward\,-Takahashi identities
(bWI).

As $S_k$ is invariant under the action of $\Der^a+\bar\Der^a$ it can
be read off from the Schwinger functional \eq{Schwinger} and
the effective action \eq{defofG}, that $\Der^a\!+\bar\Der^a$ leaves
the functional $\Gamma_k$ invariant for generic $A$ and $\bar A$
configurations, 
\begin{eqnarray}
\left(\Der^a\!+\,\bar\Der^a\right)\Gamma_k = 0. 
\label{mWI+bWIex}
\end{eqnarray}
Therefore by requiring $A\!=\!\bar A$, $\hat\Gamma_k$ is also invariant,
$\Der^a \hat\Gamma_k = 0$,
which for $k=0$ expresses the desired physical gauge invariance.  
Consequently, for $k\neq 0$, physical gauge invariance 
is encoded in the behaviour of $\Gamma_k$ under the transformation $\Der^a$. 
I emphasise again that, in order to attain this crucial result, I had to
keep track of the effects from the transformations $\Der^a$ and
$\bar\Der^a$ on $\Gamma_k$ separately.

For pure Yang-Mills theories the mWI is given by
\begin{eqnarray}
\Der^a(x)\,\Gamma_k =
\Der^a(x)\,\left(S_{\rm gf} + S_{\rm gh}\right) +
\Loop_k^a(x) + \Loop_{R,k}^a(x)\ .
\label{mWIex}
\end{eqnarray}
Both $\Loop_k$ and $\Loop_{R,k}$ display loop terms.
The first term $\Loop_k$ stands for the
well-known loop contributions to Ward\,-Takahashi identities in non-Abelian 
gauge theories originating from $\langle \Der^a (S_{\rm gf}+S_{\rm
gh})\rangle_J$, whilst the second term is due to the regulator terms
and clearly vanishes when $k\to0$. 
It follows that the mWI \eq{mWIex} turns into the standard WI for $k=0$,
\begin{eqnarray}
\Der^a\Gamma = \Loop^a_0\ .
\label{WI}\end{eqnarray}
As for the bWI, by applying $\bar\Der^a$ to
$W_k[J,\eta,\bar\eta;\bar A]$ it follows from \eq{Schwinger} and \eq{defofG} 
after some manipulations that the effective action $\Gamma_k$ obeys the 
equation 
\begin{eqnarray}
 \bar\Der^a\Gamma_k = 
\bar\Der^a(S_{\rm gf} + S_{\rm gh}) - (\Loop^a_k + \Loop^a_{R,k})\ .
\label{bWIex}\end{eqnarray}
The combined gauge invariance of $\Gamma_k$, Eq.~(\ref{mWI+bWIex}),
follows immediately from this identity and the
mWI given by~(\ref{mWIex}).\step

\noindent
{\sf Symmetries of the flow and physical gauge invariance\,:}
The implementation of a coarse-graining modifies the gauge
symmetry of the theory as I mentioned before.
At the formal level it is clear that the original 
symmetry is restored when the coarse-graining scale is removed.
A more delicate problem is to guarantee 
that this also happens at the level of the solution to the flow equation.

To understand how gauge invariance is encoded throughout the flow, 
it is pivotal to also study the action of the symmetry transformations 
on $\partial_t\Gamma_k$, Eq,\,\eq{flow}. Under the combined gauge
transformation the flow of $\Gamma_k$ transforms as
\begin{eqnarray}
(\Der^a\!+\,\bar\Der^a) \partial_t\Gamma_k=0\ . 
\label{d_Bflow}\end{eqnarray}
An immediate consequence is that $\Der^a\partial_t\hat\Gamma_k=0$. 
Note that the only input for \eq{d_Bflow} was the invariance of $\Gamma_k$. 
Thus, when the initial effective action $\Gamma_{\Lambda}$ is invariant under
$\Der^a\!+\bar\Der^a$,
it follows that the full effective action $\Ga_0$ is also invariant, 
$(\Der^a\!+\bar\Der^a)\Ga_0=0$.
In other words, \eq{mWI+bWIex} and \eq{d_Bflow} are 
the proof that the flow and the combined transformation commute. Moreover, 
$\Gamma_0$ satisfies the usual WI \eq{WI}. 
This means that the line of arguments for the background 
field formalism can be followed here as well. Hence  
the equation $\Der^a \hat\Gamma_0=0$  displays physical
gauge invariance.

Now I wish to make a final remark on the consistency of the mWI 
\eq{mWIex} with the flow. Here, as in other formulations of Wilsonian
flows in gauge theories,\,\cite{Ellwanger:1994iz,Litim:1998nf}
the flow of the mWI is proportional to itself. Such common property
states that if the effective action $\Gamma_k$ satisfies
the mWI at some scale $k$, \textit{e.g}\, the initial one $k=\Lambda$,
then $\Gamma_k$ {\it automatically} satisfies the mWI at any scale
$k$, provided it is obtained from integrating the flow equation, and
in particular, $\Gamma_0$ satisfies the usual Ward\,-Takahashi identity.\step

\noindent
{\sf Summary of the Talk\,:} 
I have established a complete set of equations relevant
for the control of gauge invariance in the ERG approach
to pure Yang-Mills theories. 
In particular, it can be inferred that there is no requirement for
additional fine-tuning conditions, despite the presence of a background
field. Moreover, I have shown that invariance of the effective action 
under a combined gauge transformation of 
all fields follows from mWI
and bWI.  More generally, two of these three properties of the 
effective action (invariance, mWI and bWI) lead to the third one. 

Consequently, a key property of the usual background field formalism 
is maintained in the ERG approach\,: by virtue of the auxiliary background 
identity it follows that {\it physical} gauge invariance
is reflected in the invariance of the effective action under the 
combined gauge transformation of all fields. 

The formalism is not only suitable for formal or analytic analysis,
but also, it elucidates the problem of how to filter the contribution
of spurious unphysical modes in a numerical computation of
approximate solutions to flow equations in gauge theories.

\vspace*{-2pt}

\section*{Acknowledgements}
FF thanks the organisers for their financial support.

\vfill
\eject
\end{document}